\documentclass[aps,prd,showpacs,nofootinbib,floats,floatfix,preprintnumbers,groupedaddress,twocolumn]{revtex4}

\usepackage{bm}
\usepackage{latexsym}
\usepackage{dcolumn}
\usepackage{amsmath,amsfonts,amssymb}
\usepackage{graphicx,epsfig}
\usepackage{amsmath}
\usepackage{fancyhdr}
\usepackage{hyperref}
\usepackage{graphicx,epstopdf}
\usepackage{physics}
\usepackage[utf8]{inputenc}

\hypersetup{
	colorlinks   = true, 
	urlcolor     = blue, 
	linkcolor    = blue, 
	citecolor   = red 
}

\begin{document}
\title{Is instability near a black hole key for ``thermalization'' of its horizon?}
\author{Bibhas Ranjan Majhi\footnote {\color{blue} bibhas.majhi@iitg.ac.in}}

\affiliation{Department of Physics, Indian Institute of Technology Guwahati, Guwahati 781039, Assam, India}

\date{\today}

\begin{abstract}
We put forward an attempt towards building a possible theoretical model to understand the observer dependent thermalization of black hole horizon. The near horizon Hamiltonian for a massless, chargeless particle is $xp$ type. This is unstable in nature and so the horizon can induce instability in a system. The particle in turn finds the horizon thermal when it interacts with it. We explicitly show this in the Schrodinger as well as in Heisenberg pictures by taking into account the time evolution of the system under this Hamiltonian. Hence we postulate that existing instability near the horizon can be one of the potential candidates for explaining the black hole thermalization.
\end{abstract}


\maketitle

\section{Introduction}
Quantum mechanically black holes (BH) are thermal objects \cite{Hawking:1974rv,Hawking:1974sw,Bekenstein:1973ur}. So far this is understood as a phenomena due to fact that the notion of uniqueness of vacuum does not comply with curved spacetimes.  Therefore one observer's vacuum may appear to be full of particles with respect to others. For instance the static observer at the asymptotic infinity finds particle in the vacuum of a freely falling observer, generally known as {\it Hawking radiation}. The horizon temperature $T$ comes out to be proportional to the BH surface gravity $\kappa$ \cite{Hawking:1974rv,Hawking:1974sw}. This event, since its prediction, has been a key to emerge as well as understand BH information paradox \cite{Mathur:2009hf} and people thinks that the Hawking evaporated particles may carry the information of infalling objects inside the BH. Investigation so far on the basis of information theory created such possibility, but additionally develops several favourable as well as counter arguments starting from complementarity proposal \cite{Susskind:1993if} to Hayden-Preskill  ``scrambling time'' \cite{Hayden:2007cs,Sekino:2008he}. Till now we are far from the goal. We feel that the failure to achieve the complete success lies in the heart of the ``{\it lack of full (theoretical) understanding  about the thermalization of horizon}''.

In this paper, we suggest that the idea of observer dependent thermalization of horizon can be illuminated in a model, build out from a massless particle moving very near to the horizon.
The Hamiltonian in near horizon regime for a massless and chargless outgoing particle, moving in static, spherically symmetric  (SSS) BH spacetime, is given by
\begin{equation}
H=\kappa xp~,
\label{1}
\end{equation}
where $x=r-r_H$ with $r_H$ is the location of the horizon and $p$ is the momentum of particle corresponding to radial coordinate $r$. This was obtained in Eddington-
Finkelstein (EF) outgoing null coordinates. In the above, the path has been chosen to be along the normal to null hypersurface, given by Eddington null coordinate $u=$ constant, where $u=t-r^*$ with $r^*$ is the well known tortoise coordinate (see \cite{Dalui:2020qpt} for details on construction of (\ref{1})). The same can also be obtained in Painleve coordinates for SSS BH as well as Kerr BH, considering a particular trajectory \cite{Dalui:2018qqv,Dalui:2019umw,Dalui:2019esx}.  Recently we explicitly showed that the trajectory of the particle feels a ``local instability'' which in turn may cause temperature to the horizon at the semi-classical level \cite{Dalui:2019esx,Dalui:2020qpt,State1}. 

Having this indication, we here investigate this model in a pure quantum mechanical way, particularly the time evolution of a wave function under (\ref{1}). We observe that the later time wave function possess thermal features with respect to a particular class of wave function which was initially absent, and $T$ is given by that of Hawking expression. 
The same has also been predicted in Heisenberg's picture as well by studying the nature of out-of-time order correlation (OTOC) between the position and momentum of the particle. Thus we feel that this particular model may has potential ability to provide hints in building a theoretical framework which will reveal the underlying physical reason (theory) to describe the ``relative thermalization'' of BH horizon.

In search of the physical reason of the above observation we find from literature \cite{BK1,BK2,MG,BK3,BGS} that the $xp$ Hamiltonian is unbounded in nature and devoid of time reversal symmetry. Moreover it can be chaotic in nature. This unbounded feature can be realised from equations of motions driven by this Hamiltonian. These are $x \sim e^{\kappa t}$ and $p\sim e^{-\kappa t}$. As in our case $t\to -\infty$ corresponds to horizon of the black hole, the momentum becomes unbounded here and consequently the motion of the particle very near to horizon  becomes unstable. This can also be realised from the following analysis as well. For a given energy $E$, the momentum of the particle is given by $p=E/(\kappa x)$ (see Eq. (\ref{1})) which diverges in limit horizon $x\to 0$. Moreover, Eq. (\ref{1}) takes the form for that of an inverted harmonic oscillator in a new canonical variables \cite{BK1,BK2,BK3} which is known to be unstable. So the unstable (or unbounded) nature is inherent to this Hamiltonian which, in our case, exits in the near horizon regime. In addition, recently it is found that such unboundedness can provide chaotic nature to a system when it interacts with the horizon (see e.g. \cite{Hashimoto:2016dfz,Dalui:2018qqv,Dalui:2019umw}). All these indicates that the horizon can provide instability in the particle motion when it is moving very near to the horizon and we feel that this local instability can be responsible for such thermalization and thereby can be a potential physical reason of such phenomenon.

We also observe that the present simple model can give estimation of various important quantities in the context of possibility of retrieving information from Hawking radiation. We find that to recover one bit of information (assuming one particle carries it) the minimum time required to be that given by Hayden and Preskill \cite{Hayden:2007cs}. If so, then each bit must contains energy which is similar to standard equipartition of energy among each degrees of freedom (DOF). We also estimate the maximum number of bits (or DOF) a BH can emit by its complete evaporation. Interestingly the out come is consistent with earlier observation \cite{Padmanabhan:2013nxa} in the context of gravity as an emergent phenomenon. Moreover, evaporation time is consistent with Page's estimation \cite{Page:1993df}. 

So in this paper a possible way to build a theory is being proposed in the context of near horizon instability to understand the thermal character of horizon.
Having this present new way of looking at the horizon thermality in a unified manner we feel that the present study and results have potential to achieve deeper understanding of this subject and may help in resolving the long standing information paradox problem, particularly in the context of information theory approach. Let us now proceed towards the calculation in support of our present claims and what we wanted to mean by thermalization.

\section{Time evolution of a state and ``thermalization''} 
In order to perform the quantum mechanical analysis of (\ref{1}), we first make it Hermitian. For this, using Weyl ordering, we express it in the following operator form:
\begin{equation}
\hat{H}=\frac{\kappa}{2}(\hat{x}\hat{p}+\hat{p}\hat{x}) = \kappa \Big(\hat{x}\hat{p} - \frac{i\hbar}{2}\Big)~,
\label{2}
\end{equation}
where in the last step the commutator $[\hat{x},\hat{p}]=i\hbar$ has been applied.
Considering the unitary time evolution of an arbitrary quantum state $\Psi(x,0)$ (representing our test particle on the stationary background), one can find the later time state as \cite{SF} {\footnote{It is easy to check that the norm of the wave function preserves with time evolution (see Appendix \ref{Norm}).}}
\begin{equation}
\Psi(x,t) = e^{-\frac{\kappa t}{2}} \Psi(x e^{-\kappa t},0)~.
\label{8}
\end{equation}
In the above $t >0$ is the time lapse between the initial state and the later state, representing our particle under study. For the choice initial time as $t_{i}=0$, $t$ will be the later time at which we are interested to know about the evolved wave function. The evolution is done under the Hamiltonian (\ref{2}) and hence the time evolution operator is given by $U=e^{-\frac{i\hat{H}t}{\hbar}}$. Here we assumed that the wave function starts evolving from time $t_i=0$ so that it has no interaction with the Hamiltonian before initial time. After that it is in interaction with the Hamiltonian. Pictorially, it can be thought as keeping the particle, represented by the wave function $\Psi(x,0)$, very near to the horizon from time $t_i=0$; so that after time $t$ the wave function evolves to $\Psi(x,t)$ as it is now under the Hamiltonian (\ref{2}).



Note that the later time quantum state suffers a dilation and this can lead to a mixing of positive and negative momenta in the evolved state, even though initial state is in particular momentum (positive or negative, not both). Due to this mixing the state becomes thermal in nature with respect to a relevant monochromatic plane wave. 
We shall now investigate this possibility for two simple one dimensional examples: (a) a plane wave and (b) a Gaussian wave packet.

(a) {\underline{Plane wave:}} The initial wave function is forward propagating plane wave $\Psi(x,0) \sim e^{ik_0x}$ with the momentum is defined as $p_0 = \hbar k_0$. Later time this will evolve according to (\ref{8}):
\begin{equation}
\Psi(x,t) \sim e^{-\frac{\kappa t}{2}} \exp[{ik_0xe^{-\kappa t}}]~.
\label{17}
\end{equation}
We shall now investigate whether the later time wave function (\ref{17}) is thermal in nature. This can be shown by calculating the power spectrum of the wave, which is determined by taking the Fourier transform of the following form \cite{Paddy1} (a justification, following \cite{Singh:2013dia,Paddybook} is being presented in Appendix \ref{App1}):
\begin{eqnarray}
f(-\omega) &\sim& \int_{-\infty}^{+\infty} dt e^{-i\omega t} e^{-\frac{\kappa t}{2}} \exp[{ik_0xe^{-\kappa t}}]
\nonumber
\\
&=& \frac{1}{\kappa\sqrt{k_0 x}} e^{i(\frac{\pi}{4}-\frac{\omega}{\kappa}\ln(k_0 x))}e^{-\frac{\pi\omega}{2\kappa}}\Gamma\Big(\frac{1}{2}+\frac{i\omega}{\kappa}\Big)~,
\label{17.1}
\end{eqnarray}
where $\omega>0$.
Utilizing the identity  $|\Gamma\Big(\frac{1}{2}+ia\Big)|^2=\pi/\cosh(\pi a)$ one finds
\begin{equation}
|f(-\omega)|^2\sim \frac{2}{\kappa^2 k_0 x}\frac{1}{e^{\frac{2\pi\omega}{\kappa}}+1}~,
\label{17.2}
\end{equation}
which is thermal in nature with the temperature is given by the Hawking expression
\begin{equation}
T=\frac{\hbar\kappa}{2\pi}~.
\label{T}
\end{equation}
In the above energy $E=\hbar\omega$ has been taken. 

It may be pointed out that expression (\ref{17.2}) is considered to be thermal (upto an over all factor) in the sense of usual Fermi distribution (further justification can be followed from \cite{Paddy1,Singh:2013dia,Paddybook}). Our obtained result is identical to Fermi distribution upto a pre-factor. The appearance of pre-factor also occurs in the original calculation by Hawking. For black hole system, in original calculation, this usually arises due to the back scattering of the emitted particles by the effective potential which exists in between the horizon and infinity. Such a modification in the {\it ideal} thermal expression in the original analysis is usually known as {\it grey-body} factor. Therefore upto this factor one takes the result as thermal. Although the present factor in Eq. (\ref{17.2}) can not be treated as grey-body factor (as our calculation is very near to the horizon and the pre-factor can be huge as $x$ is very small, although not exactly equal to zero), but using the sprit of original analysis we call (\ref{17.2}) as thermal. In later part of this section and in the concluding section we will again comment on the realization of thermalization from (\ref{17.2}).

Of course, the presence of pre-factor in (\ref{17.2}), in true sense, signifies that the expression is not pure thermal one. In this situation a better way of looking at the result is to find the ratio $|f(-\omega)|^2/|f(\omega)|^2$ which is reminiscent to the ratio between the Bogoluibov coefficients (a similar interpretation has been adopted in Section $4.2$ of \cite{Singh:2013dia} in the case of thermal properties of de-Sitter horizon through the mixing coefficient). Then the present result yields the ratio as
\begin{equation}
\frac{|f(-\omega)|^2}{|f(\omega)|^2} = e^{-\frac{2\pi\omega}{\kappa}}~,
\label{New1}
\end{equation}
which is identical to the Boltzmann factor. Therefore one again identifies the temperature as (\ref{T}). In this way of viewing the result does not depend on the pre-factor as it cancels and thereby providing a more elegant explanation of thermalization.  

\begin{figure}[!ht]
	\centering
	\includegraphics[scale=0.50, angle=0]{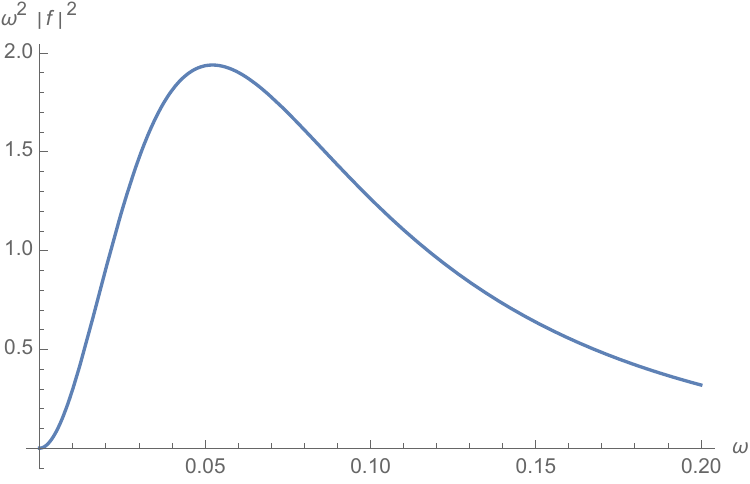}
	\caption{(Color online) $\omega^2|f(-\omega)|^2$ Vs $\omega$ plot for $\kappa=0.1$, $x=0.1$ and $ d=1=k_0$.}
	\label{Fig3}
\end{figure}

(b) {\underline{Gaussian wave packet:}}
Consider the initial wave packet propagating along positive $x$-direction as
\begin{equation}
\Psi(x,0) \sim e^{ik_0x- \frac{x^2}{2d^2}}~.
\label{19}
\end{equation}
Then at time $t$ it will evolve to
\begin{equation}
\Psi(x,t) \sim e^{-\frac{\kappa t}{2}}\exp[{ik_0xe^{-\kappa t} - \frac{x^2e^{-2\kappa t}}{2d^2}}]~.
\label{20}
\end{equation}
Fourier transformation of (\ref{20}) yields
\begin{eqnarray}
&&f(-\omega) 
= \frac{1}{\kappa}2^{-\frac{3}{4}+\frac{i \omega}{2 \kappa}} \left(\frac{x^2}{d^2}\right)^{-\frac{3}{4}-\frac{i w}{2 a}}
\nonumber
\\
&\times& \Big\{\frac{x}{d}\Gamma\Big(\frac{\kappa+2i\omega}{4\kappa}\Big)  {_1}F_1\Big[\frac{\kappa+2i\omega}{4\kappa};\frac{1}{2};-\frac{d^2k_0^2}{2}\Big]
\nonumber
\\
&+&i\sqrt{2}k_0x\Gamma\Big(\frac{3}{4}+\frac{i\omega}{2\kappa}\Big)    {_1}F_1\Big[\frac{3}{4}+\frac{i\omega}{2\kappa};\frac{3}{2};-\frac{d^2k_0^2}{2}\Big]\Big\}~,
\label{24}
\end{eqnarray}
where $_1F_1[a;b;c]$ is the Kummer confluent hypergeometric function. Like earlier, one needs to compute $|f(-\omega)|^2$ from the above expression. But this will not provide any suitable analytical form to investigate. Therefore whether  the above one corresponds to thermal distribution, we plot $\omega^2 |f(-\omega)|^2$ as a function of $\omega$ in Fig. \ref{Fig3}.
It is found that the distribution is thermal in nature in the sense that the feature is exactly similar to usual Fermi distribution (more precisely to  $\omega^2$ times the Fermi occupation number). An analytical expression for temperature is not possible to obtain here. But a numerical comparison between these two must yields the required expression. In this scenario we at least observed that the later time wave function occupies thermal property with respect to monochromatic waves like $e^{-i\omega t}$. It confirms that the time evolved wave packet turns out to be thermal with respect to plane monochromatic wave, although it was not initially.

In the above discussion it must be noted that the later time wave functions (\ref{17}) and (\ref{20}) have been constructed by using unitary time evolution under the Hamiltonian (\ref{1}). Surprisingly the final wave functions become thermal, even if the initial is not. This statement needs to be taken carefully. Note that since the initial state is not a mixed state and the time evaluation is unitary, the later wave function does not provide a mixed state. Therefore this latter wave function is not thermal {\it with respect to its own observer}. Here ``thermalization'' word has been used with respect to a different class of observers. Note that the initial wave function (i.e. $t=0$) (for simplicity considering the first example) with respect to an observer who carries a monochromatic wave of the form $e^{-i\omega t}$ will appear as
\begin{equation}
f(-\omega) \sim \int_{-\infty}^{+\infty} dt e^{-i\omega t }e^{ik_0 x_0} \sim e^{ik_0x_0} \delta(\omega)~,
\label{EPJC1}
\end{equation}
which vanishes as $\omega>0$. Similar also applies for the other example as well. Therefore there is no mixing coefficient.
Whereas the later time wave function leads to nonvanishing mixing coefficient with respect to $e^{-i\omega t}$ which appears to be thermal in character. Hence at later time the evolved wave appears to be thermal with respect to this particular observer who carries this monochromatic wave; although the same was not at $t=0$. Such an observer dependent thermal character is quite consistent with the existing concept of black hole thermodynamics. In the latter case the horizon is thermal with respect to certain class of observers -- e.g. the static observer sees the horizon as thermal object while an observer, freely falling towards the horizon,  will find that such is not true. In the whole paper this notion of concept as ``thermalization'' is being adopted. 
 Since the time evolution is done by the unitary transformation, we feel that such is completely due to the nature of our particular Hamiltonian (\ref{1}) which has been provided by the horizon in its vicinity. Hence we state that {\it the horizon shows its presence by thermalizing a quantum system and in reverse makes the system to feel BH as an observer dependent thermal object}. All these are with respect to a particular observer (here it is with respect to the monochromatic wave). We feel that this is happening due to the unstable nature of the driving Hamiltonian. Therefore deeper understanding of such instability may help us to know more about the reason for thermal behaviour of horizon.
 A further comment on this concept thermalization is as follows. Here a very simple model has been adopted and the thermalization is tested indirectly through the motion of a particle very near to the horizon. We found that the instability provided by the horizon is reflected in the observer dependent thermalization at the quantum level. A direct way of understanding this has to be the study of quantum nature of horizon itself; not by the features of auxiliary particle in the spacetime. Presumably a black hole is effectively a many body system and study of such many body effective system should lead to a direct realization of thermalization. Such is beyond the scope of this study. But the idea of emergence of instability by the horizon might be important to illuminate the effective many body theory in the context of black hole. Therefore the present single particle features can be helpful for the actual goal. This is somewhat suggestive, but not a conclusive one.

We just saw that the later time wave function yields non-vanishing Bogoliubov coefficient like quantity $f(-\omega)$ (called as mixing coefficient) which quantifies the mixing of positive and negative frequencies (or momentum here) instantaneous monochromatic plane waves (see Appendix \ref{App1} for justification of this comment). In this sense we call this thermalization mechanism as {\it relative thermalization} with respect to plane monochromatic waves. A more direct evidence for demanding the present one as a case of relative thermalization is being given in Appendix \ref{App2}. Here by calculating Bogoluibov coefficients between the eigenstates of Hamiltonian for a free particle (e.g. in absence of gravity which can be chosen in the asymptotically infinite radial distance) and those of  Hamiltonian  (\ref{1}) we show that the eigenstates of $\kappa xp$ Hamiltonian appear to be thermal in nature with respect to the other one. Later we will again make comments on the issue of thermalization.


\section{Thermalization from ``horizon as a scrambler''}
Now we will show that the similar thermalization can be addressed at the operator evolution level.
The OTOC $A(t)=\bra{}\Big|[\hat{x}(t),\hat{p}(0)]\Big|^2\ket{}$ can be evaluated as follows. $\hat{x}(t)$ is obtained by using Heisenberg's equation of motion:
\begin{equation}
\frac{d\hat{x}(t)}{dt} = \frac{1}{i\hbar}[\hat{x}(t),{\hat{H}}]~.
\label{27}
\end{equation}
Using the form for $\hat{H}$ from the first equality of (\ref{2}) and the commutator $[\hat{x},\hat{p}]=i\hbar$ in (\ref{27}) one finds
$\hat{x}(t) = \hat{x}(0)e^{\kappa t}$.
Hence the OTOC is given by
\begin{equation}
A(t) = \hbar^2 e^{2\kappa t}~,
\label{29}
\end{equation}
where we have used the fact that $\ket{}$ is a normalised state. Note that OTOC increases exponentially with time. Since $xp$ Hamiltonian represents inverted harmonic oscillator (IHO) in a new set of canonical variables, the above result can also be extracted from OTOC calculated in \cite{Bhattacharyya:2020art} for IHO. This shows that BH horizon acts as scrambler and the sensitivity is controlled by Lyapunov exponent $\lambda_L=\kappa$. In this context it may be mentioned that in a many body system the upper limit of $\lambda_L$ is controlled by the system's temperature \cite{Maldacena:2015waa}: $\lambda_L\leq (2\pi T)/\hbar$. In this case the dual BH geometry incorporates temperature, given by the saturation value. This led to conjecture that the system with positive $\lambda_L$ incorporates temperature in the semi-classical regime \cite{Morita:2019bfr}. Therefore our above analysis provides the concept of temperature which is again given by (\ref{T}). It shows that the horizon acts as scrambler on the particle motion and consequently generates temperature at the quantum (semi-classical) level. We feel that such feature of horizon is responsible to perceive BH as a thermal object.

This scrambling leads to the distribution of the system information among all possible eigenstates of it. After certain time scale, known as scrambling time, the distribution will be such that the average value of any observable is always near to the thermal equilibrium (microcanonical ensamble average) value. This sometimes known as quantum ergodicity. This is the main reason why the horizon incorporates thermality. We will shortly discuss this more elaborately. The time scale is determined by setting $A(t)\sim\mathcal{O}(1)$. From (\ref{29}) we find the scrambling time as 
\begin{equation}
t_* \sim \Big|\frac{1}{2\kappa}\ln\frac{1}{\hbar^2}\Big|=  \Big|\frac{1}{\kappa}\ln\hbar\Big|~.
\label{33}
\end{equation}
This is sometime called as Ehrenfest time. Classical to quantum correspondence principle tells that $t_*\rightarrow\infty$ as $\hbar\rightarrow 0$. This is because to obtain classical ergodicity one needs to wait for sufficiently long time. This is exactly happening for the above one as well and hence our predicted scrambling time is consistent with the well celebrated correspondence principle. Now for Schwarzschild BH we have $\kappa=1/(4M)$, where $M$ is the mass of BH and so $t_*\sim M\ln\hbar$. The above can also be expressed in terms of $T$ as $t_* \sim \frac{\hbar}{2\pi T}\ln\hbar$.

So far we observed that relative thermalization of horizon is occurring due to evolution of system under the Hamiltonian (\ref{1}) (either in Schrodinger or in Heisenberg picture). Such is due to the spacial characteristic feature is provided by the Hamiltonian in the near horizon regime. As explained elaborately in the Introduction, the unstable character of the Hamiltonian at the classical level might be the main reason. Therefore we feel that the classical instability near the horizon is reflected through the apparent thermalization of the horizon at the quantum regime.
 It may be pointed out that the initial wave function represents particles of momentum $p_0$ whose value can be anything, say $0$ to $N$. But later time the momentum gets red shifted $p(t)=p_0e^{-\kappa t}$ and hence its possible values must cover a smaller range which is from $0$ to $m<N$. Therefore at later time, the particles, represented by this wave function, cover subspace of the initial momentum space. On other side at $t=0$, $\Psi(0,x)$ is extended for all $x$. Whereas since (\ref{1}) is valid only for $x>0$, the later wave function $\Psi(x,t)$ confined in $x>0$ region. In this sense it covers less ``DOF'' compared to its actual size at $t=0$. Moreover, the probability density $|\Psi(x,t)|^2\sim e^{-\kappa t}$, which can be interpreted as diagonal element of density matrix in position basis (i.e. $\rho_{xx} = \bra{x}\ket{\Psi_t}\bra{\Psi_t}\ket{x}$), is suppressed for $t\neq 0$ and leading to more and more ``loss of information'' for this particular eigenbasis as $t$ increases. Moreover here the whole system is composed of two objects: BH and particle; and they are interacting through (\ref{1}). Note that the particle wave function (e.g. the plane wave here) is evolved under this Hamiltonian, not by its own free particle Hamiltonian. Additionally, the particle is not isolated from the environment (like black hole here) as it is in constant interaction with BH.  This BH provides the local instability in the particle's trajectory and in turn at the quantum level the particle finds the horizon as thermal object. This explains the physical reason why both levels of  evolution -- wave function as well as operator -- leads to (relative) thermalization and thereby providing a possible strong mechanism for well known observer dependent thermal behaviour of horizon. 
 
 Before moving forward we want to show that the eigenstates of our Hamiltonian (\ref{1}) are indeed thermal in nature with respect to the arbitrary state of a system. Choosing the basis as eigenstates of (\ref{1}), an arbitrary state can be expanded as $\ket{\Phi} = \sum_E c_E\ket{E}$. Here $|c_E|^2 = |\bra{E}\ket{\Phi}|^2$ is the probability of finding $\ket{\Phi}$ in eigenstate $\ket{E}$ with energy $E$. Using position kets $\ket{x}$ as basis one finds
\begin{equation}
c_E = \int_0^\infty dx \Phi(x) u_E^*(x)~,
\label{CE}
\end{equation}
where $u_E(x)$ is the normalized eigenfunction of (\ref{1}) in position representation, which is given by (see Section VII of \cite{Dalui:2020qpt})
\begin{equation}
u_E(x) = \frac{1}{\sqrt{2\pi\hbar\kappa}}(x)^{-\frac{1}{2}+\frac{iE}{\hbar\kappa}}~.
\label{UE}
\end{equation}
To perform the integration in (\ref{CE}) analytically, we choose a simple test wave function $\Phi(x)\sim e^{-ik_0x}$. Using all these in (\ref{CE}) one easily obtains
\begin{equation}
|c_E|^2 = \frac{1}{\hbar\kappa k_0} \frac{1}{e^{\frac{2\pi E}{\hbar\kappa}} + 1}~.
\label{B123}
\end{equation}
Now for the ideal Fermi particles we know that the probability of finding $n$ number of particles in state $E$ is given by (see section $6.3$ of \cite{Book})
\begin{eqnarray}
P_E(n) = \begin{cases}
1-<n_E>,~~~ \textrm{for}~ n=0\\
<n_E>, ~~~\textrm{for}~ n=1~,
\end{cases}
\end{eqnarray}
where 
\begin{equation}
<n_E> = \frac{1}{e^{\frac{E}{T}}+1}~.
\end{equation}
This shows that the probability distribution of each eigenstates of our Hamiltonian (\ref{1}) with respect to the test wave function is thermal (upto to a pre-factor) if one compares with the probability of finding $n=1$ number of particles in energy state $E$ for Fermions. 


Below we will show that various significant quantities, related to the possibility of information retrieval from Hawking radiation, can be estimated through this simple model.

\section{On the information retrieving from Hawking radiation}
The general picture of Hawking radiation is described in the following way. Near the horizon there are always particle-antiparticle pairs. One of the member of this pairs, namely anti-particle, remains within BH while the other one (the particle) escapes from the horizon barrier to reach at infinity. Escape through the horizon barrier is possible only at quantum level and so the Hawking radiation is a quantum phenomenon. Here we consider that 
{\it Hawking radiation from BH occurs only when the particle and antiparticle, within a pair, are at least Planck length $l_p$ order distant apart.} Here $l_p$ is taken as it is the possible minimum physical distance.  
By this we meant the antiparticle remains within the horizon and the particle is visible as Hawking radiated one when it travels at least $l_p$ from the horizon in radial direction. Below this, it is still a part of BH. 

Now just after coming out of the horizon, particle will see the horizon and so its motion will be dominated by Hamiltonian (\ref{1}) if it is massless and chargeless (like photon) and as long as it is very near to the horizon. In this case the radial trajectory is given by $x=\frac{1}{\kappa}e^{\kappa t}$ \cite{Dalui:2020qpt,Dalui:2018qqv,Dalui:2019umw,Dalui:2019esx}.
Note that, as stated earlier, it has been obtained in EF and Painleve coordinates. The SSS BH metric in EF is given by (see \cite{Dalui:2020qpt} for details)
\begin{eqnarray}
ds^2 &=& -f(r) dt^2+2(1-f(r))dtdr
\nonumber 
\\
&+& (2-f(r))dr^2 +r^2(d\theta^2+\sin^2\theta d\phi^2)~.
\end{eqnarray}
$f(r_H)=0$ determines the horizon location and near to this we have $f(r) \simeq 2\kappa x$.
Then the physical radial distance is given by 
$r_{phy} = \int_{r_H}^r dr\sqrt{2-f(r)}$,
which at the leading order in the near horizon regime yields
\begin{eqnarray}
r_{phy} \sim \int_{0}^{x}dx(1-\frac{\kappa x}{2})\simeq x~.
\label{RadialPhy}
\end{eqnarray}
Similarly in Painleve also the physical radial distance is given by (\ref{RadialPhy}). Therefore we conclude that the nature of the physical radial distance, in the near horizon regime, characterized by $x=\frac{1}{\kappa}e^{\kappa t}$.
 With this we now want to calculate the minimum time taken by the particle to be identified as Hawking particle assuming that the dynamics is driven by (\ref{1}), even within distance $l_p$. Since the minimum distance needs to be $x= l_p$, estimated the minimum time is then given by 
\begin{equation}
t_{\textrm{min}} \sim \frac{1}{\kappa}\ln(l_p\kappa)~.
\label{34}
\end{equation}
We may call this as the minimum time one needs to spend near the horizon to get one bit of information in the form of massless particle as Hawking radiation.
Interestingly, for Schwarzschild BH we have $|t_{\textrm{min}}| \sim M\ln (M/l_p)$ which is identical to Hayden-Preskill time \cite{Hayden:2007cs}.

Now let us estimate how much minimum energy the observer will register within time $t_{\textrm{min}}$ who is detecting the Hawking particle. Uncertainty relation between energy and time will give this estimation. For each particle this will be given by
\begin{equation}
E_{\textrm{min}} \sim \frac{\hbar}{t_{\textrm{min}}} = \frac{\hbar\kappa}{\ln(l_p\kappa)} = \frac{2\pi T}{\ln(l_p\kappa)}~.
\label{35}
\end{equation}
It implies that each radiated particle carries above minimum amount of energy or in other words, at least this energy is required to create one Hawking particle. In this situation if we assume that the whole BH of energy $\mathcal{E}$ is radiated by this type of particle, then the maximum number emitted particles will be
\begin{equation}
N_{\textrm{max}}\sim \frac{\mathcal{E}}{E_{\textrm{min}}} = \frac{\mathcal{E}\ln(l_p\kappa)}{2\pi T}~.
\label{36}
\end{equation} 
Next use of Gibbs-Duhem like relation $2TS=\mathcal{E}$ for BH, known as Smarr formula \cite{Smarr:1972kt,Padmanabhan:2003pk,Banerjee:2010yd,Banerjee:2010ye}, yields
\begin{equation}
N_{\textrm{max}} \sim \frac{S\ln(l_p\kappa)}{\pi}~,
\label{37}
\end{equation}
which implies that the maximum number of emitted particles is proportional to BH entropy. 

Now we postulate that 
{\it each emitted particle contains one bit information of BH.}
Then the estimated maximum number of bits, representing BH, is given by (\ref{37}) and therefore each bit contains energy, given by (\ref{35}). Note that (\ref{37}) is proportional to BH entropy which depends on horizon area in Einstein's gravity. It indicates that all the information of BH are contained on the surface of it. Interestingly, the above one is consistent with an earlier result, given by Padmanabhan \cite{Padmanabhan:2013nxa}, to estimate the number of surface DOF of a BH based on equipartition of energy. Then time for complete evaporation is $t_{\textrm{min}}N_{\textrm{max}}\sim (S/\kappa)(\ln(l_p\kappa))^2/\pi$ which is for Schwarzschild BH turns out to be $\sim M^3$. Note that this time scale is of the order that predicted by Page \cite{Page:1993df}.

We conclude this discussion by mentioning the recent developments in Page curve in the context of information paradox problem.  Page curve \cite{Page:1993wv,Page:2013dx} suggests that for the unitarity, initially the entanglement entropy should be equal to Hawking radiation and must increase till a certain time. At this time, known as ``Page time'', both black hole and radiation entropies become equal. After that entanglement entropy decreases and always equal to black hole entropy. The reason for this nature of Page curve has been recently elaborately described in \cite{Penington:2019npb,Almheiri:2019psf} in the context of gravitational point of view using Ryu - Takayanagi (RT) formula \cite{Ryu:2006bv}. In early time the RT surface includes all the interior to form pure state and the early Hawking radiation leads to increase in entanglement entropy. Later time this surface moves near to the horizon and providing the decrease in entanglement entropy. In this portion, i.e. after Page time, to reconstruct one bit information which was thrown to black hole one needs to wait a time given by scrambling time as predicted earlier. The present discussion, although being based on very basic semi-classical approach, is capable of reproducing several related old results and moreover makes a connection to them with earlier results, done in a completely different angle, like given in (\ref{36}) and (\ref{37}). The positive side of our model is it not only provides a way of understanding the horizon thermalization, but also goes beyond that. It may be noted that the number of bits, as estimated in (\ref{37}), depends on black hole entropy. Therefore we feel that the results obtained in this section, in the light of Page curve, are much more suitable after Page time regime.

\section{Conclusions}
In this paper we showed that the time evolved system appeared to be thermal under the $xp$ type Hamiltonian with respect to certain waves (sates). Since the near horizon massless, chargeless particle follows this model, we suggest that the phenomena of horizon (relative) thermalization probably can be illuminated through the knowledge provided by this model. The temperature is derived to be that of the Hawking expression. We discussed that since $xp$ Hamiltonian induces instability in the particle motion,
probably the horizon thermalization is a particular case of existing instability in the near horizon region, thereby we are providing a possible distinct idea to understand the underlying physical reason for perceiving BH as a thermal object. Note that the obtained spectrum (\ref{17.2}) is Fermionic in nature. This may be due to the spacial form of our Hamiltonian which contains first order derivative with respect to $x$ as one writes the Schrodinger type equation for $H\sim xp$. Such character is very similar to Dirac equation for spin half particle. But it may be speculative and hence to get conclusive statement further understanding about the $xp$ Hamiltonian is necessary.


Our Hamiltonian (\ref{1}) represents a single-particle and integrable system. Therefore, although (\ref{1}) is unstable in nature, the obtained thermalization can not be described by  Gibbs ensemble (GE). In this regard it may be pointed out that a possible connected mechanism can be based on generalized Gibbs ensemble (GGE). GGE is well applicable for an integrable few-body system, but does not describe a genuine thermalization. But whether the present thermalization can be described by GGE is still illusive at this moment. It needs further investigation and analysis to find a concrete answer in this direction. On the other hand, the thermalization in the present case is being determined by the existing idea for the black holes. Till date this is being investigated by calculating the Bogoluibov coefficients between mode functions for the two observers. It is found that one observer's positive frequency mode appears to be mixture of positive and negative frequency modes of the other observer. In that case the modulus square of Bogoluibov coefficient $\beta$ takes the form of usual thermal distribution (Bose-Einstein for scalar modes and Fermionic for Dirac mode) upto a grey-body factor. In that case comparison between $|\beta|^2$ with usual one concludes that one observer's mode is thermal with respect to other one's and therefore the thermalization of horizon is an observer dependent phenomenon. In the present analysis identical idea has been adopted. Therefore as explained in the main text, we call this thermalization as {\it relative thermalization}. Mainly the time evolved wave function by the  Hamiltonian (\ref{1}) or its eigenstates appear to be thermal in nature with respect to certain wave function. Now since (\ref{1}) unstable in nature, we feel that this instability is responsible for such mixing and thereby yields relative thermalization. But whether it is a genuine thermalization in the sense of GE is still an open question. We are working in this direction and as far as we  know, till date it is not well understood. Hope, on the basis of this analysis, the present model will be able to enlighten this direction.

In addition, within this framework, we also showed that various relevant quantities can be estimated which are related to the subject of the possibility of information collection from the Hawking radiated particles. Although the analysis is based on certain assumptions, but this heuristic approach is capable of giving important information of BH radiation which are the subject of investigation within the information theory approach \cite{Hayden:2007cs}. Our estimated time to collect one bit of information for an outside observer, very close to the horizon, nicely matches with that predicted by Hayden and Preskill \cite{Hayden:2007cs,Sekino:2008he}. Moreover, the minimum energy of each bit is quite consistent with the equipartition of energy (see Eq. (\ref{35})). Finally the estimated number of bits in the complete evaporation of a BH came out to be very much identical to the predicted one in literature \cite{Padmanabhan:2013nxa}, while the total evaporation time scale matches with Page's  estimation \cite{Page:1993df}.

In this analysis we observed that the present simple model can suggest the thermal properties of black hole. The analysis does support the observer dependent thermal nature of horizon which was initially suggested by Hawking through quantum field theory. Although such prediction is not new except the simplicity of the current model, but there are few new features in the present model. First of all we found that the driving Hamiltonian is unstable in nature. Therefore, even if it was known that the particular set of observers sees thermal nature because of mixing of positive and negative energy modes, but we now know that such mixing is happening due to the presence of instability near the horizon. So our model probably can revel the physical mechanism for such mixing. Moreover this model has a proposal for the governing Hamiltonian which is responsible for such event. Such knowledge may have extra benefit in understanding the topic of investigation. Secondly, apart from Hawking's original calculation people are trying to understand the Hawking effect and related black hole thermodynamics in various approaches with the hope that these will revel more about them. The present way of investigation can be one of such attempts. Finally, we found that within this model various important quantities related to information collection through Hawking radiation can be estimated. Hence the present attempt can have some degree of usefulness in understanding thermal behaviour of horizon and information retrieve issues. But this suggestion needs further investigation in order to achieve any conclusive direction.

We now conclude with the statement that we here, within the $xp$ Hamiltonian model, found a possible direction to build the underlying physical theory which can explain the thermalization of horizon. This is the first attempt, so far as we know, to explain such phenomenon within in a concrete theoretical background. Moreover, we showed that this model is very effective in explaining various important information about the BH evaporation in a unified way. Hope this present findings will help to understand this paradigm more in future.

\vskip 5mm
{\bf Acknowledgements}: {\sc This work is dedicated to those who are helping us to fight against COVID-19 across the globe.}

The author thanks Surojit Dalui, Sumit Dey and Partha Nandi for helpful suggestions and comments during the progress of the work. 
This work is inspired by the works of Prof. T. Padmanabhan and, I dedicate this paper to his memory.



\vskip 3mm
\noindent
{\bf Data availability:} This manuscript has no associated data or the data will not be deposited.


\newpage
\begin{widetext}
\begin{center}	
\end{center}
\appendix
\section{Norm preserves under time evolution}\label{Norm}
We first calculate the evolution of a wave function, representing the state of our test particle. The propagator for evolution at $(t_i=0, x_1)$ to $(t_f=t, x_2)$ is given by (follow the procedure as done in \cite{SF})
\begin{equation}
G(x_2,t;x_1,0) = \bra{x_2}U\ket{x_1} = e^{-\frac{\kappa t}{2}}\delta (x_1 - x_2e^{-\kappa t})~,
\label{N1}
\end{equation}
with $t>0$ and $U = e^{-\frac{i\hat{H}t}{\hbar}}$.
Then the wave function at time $t$ is obtained as
\begin{eqnarray}
\Psi(x,t) &=& \bra{x}\ket{\Psi(t)} =  \bra{x}U\ket{\Psi(0)} = \int dx'  \bra{x}U\ket{x'}\bra{x'}\ket{\Psi(0)} =\int dx' e^{-\frac{\kappa t}{2}}\delta (x' - xe^{-\kappa t}) \Psi(x',0)
\nonumber
\\
&=& e^{-\frac{\kappa t}{2}}\Psi(xe^{-\kappa t})~.
\label{N2}
\end{eqnarray} 
In the above $\ket{\Psi(0)}$ represents the state ket at initial time $t_i=0$ and $\ket{x}$ is the eigenket of position operator.
On the other hand the complex conjugate of the above is computed as follows:
\begin{eqnarray}
\Psi^*(x,t) &=& \bra{\Psi(t)}\ket{x} =  \bra{\Psi(0)}U^\dagger\ket{x} = \int dx'  \bra{\Psi(0)}\ket{x'}\bra{x'}U^\dagger\ket{x} =\int dx' e^{\frac{\kappa t}{2}}\delta (x - x'e^{\kappa t}) \Psi^*(x',0)~.
\label{N3}
\end{eqnarray} 
In the above we have used the relation $\bra{x'}U^\dagger\ket{x} = e^{\frac{\kappa t}{2}}\delta (x - x'e^{\kappa t})$. This is because $\bra{x'}U^\dagger\ket{x} = \bra{x'}e^{\frac{i\hat{H}t}{\hbar}}\ket{x}$ can be obtained from (\ref{N1}) with $t\to - t$.
Then the norm is given by
\begin{eqnarray}
\int dx \Psi^*(x,t) \Psi(x,t) &=& \int dx dx' dx'' e^{\frac{\kappa t}{2}}\delta (x - x'e^{\kappa t}) \Psi^*(x',0) e^{-\frac{\kappa t}{2}}\delta (x'' - xe^{-\kappa t}) \Psi(x'',0)
\nonumber
\\
&=&\int dx dx' dx'' \delta (x - x'e^{\kappa t}) \Psi^*(x',0) \delta (x'' - xe^{-\kappa t}) \Psi(x'',0)
\nonumber
\\
&=& \int dx dx' \delta (x - x'e^{\kappa t}) \Psi^*(x',0)  \Psi(xe^{-\kappa t},0)
\nonumber
\\
&=& \int dx' \Psi^*(x',0)  \Psi(x',0)~.
\label{N4}
\end{eqnarray}
In third equality we have performed the integration over $x''$ and finally the integration over $x$ has been done. This analysis shows that the norm does not change under time evolution, which is consistent with the property of the unitary evolution of the wave function. 

\section{Mixing of positive and negative frequencies}\label{App1}
This analysis has a similarity with Bogolyubov coefficient calculation. For details please follow Ref. \cite{Singh:2013dia} (also see section $14.3$ of \cite{Paddybook}). Since every details are there the reader may be requested to go through these refs. These types of analysis, related to black holes, are very well known in literature.
A brief analysis is as follows.
If the later time wave function can be expressed as mixture of positive and negative frequencies, then we must have
\begin{equation}
\Psi(t) = \int_0^\infty \frac{d\omega}{2\pi} \Big[\alpha_\omega e^{-i\omega t} + \beta_\omega e^{i\omega t}\Big]~,
\label{R1}
\end{equation}
with $\omega>0$. The modes $e^{\pm i\omega t}$ are interpreted as instantaneous monochromatic plane waves in a comoving frame with respect to time $t$. This can also be expressed as (using $\omega\to -\omega$ in the second term)
\begin{equation}
\Psi(t) = \int_{-\infty}^{\infty}\frac{d\omega}{2\pi}f(\omega)e^{-i\omega t}~,
\label{R2}
\end{equation}
with the identifications $\alpha_\omega = f(\omega)$ and $\beta_\omega = f(-\omega)$. These $\alpha_\omega$ and $\beta_\omega$ are known as mixing coefficients. These are similar to Bogoliubov coefficients, but not same.  
Now in the light of (\ref{R2}) we have
\begin{eqnarray}
\alpha_\omega = f(\omega) = \int_{-\infty}^\infty dt \Psi(t) e^{i\omega t}~;
\label{R3}
\\
\beta_\omega = f(-\omega) = \int_{-\infty}^\infty dt \Psi(t) e^{-i\omega t}~;
\label{R4}
\end{eqnarray}
Note that if $\beta_\omega$ vanishes, then there is not mixing. Otherwise the mixing of both positive and negative frequency modes is there and in the light of Bogoliubov analysis $|\beta_\omega|^2$ can be treated as power spectrum. In our paper we exactly calculated this  in Eqs. (\ref{17.2}) and (\ref{24}) and found that the spectrum is thermal in nature, just like what happened in Eq. (39) of Ref. \cite{Singh:2013dia}.  Under this logic we argue that the later wave function, with respect to instantaneous monochromatic plane wave,  possess thermal nature. Therefore we call this as {\it relative thermalization}.

\section{Relative thermalization: non-vanishing Bogoliubov coefficient $\beta$}\label{App2}
The Hamiltonian for a relativistic outgoing massless particle, in absence of black hole (i.e. particle is free), is given by $H_f=p$ and so its outgoing eigenstate is of the form $\Psi_{k_0} = N e^{ik_0 x}$, where $k_0 = p/\hbar$. Here $N$ is the normalization constant.
On the other hand the eigenstates of Hamiltonian (\ref{1}) are given by (\ref{UE}). The eigenstate $\Psi_{k_0}$ can now be expressed as a linear combination of of the eigenstates (\ref{UE}) in the following form:
\begin{equation}
\Psi_{k_0} (x) = \sum_E \Big(\alpha_{k_0E}u_E(x) + \beta_{k_0E}u_E^*(x)\Big)~,
\label{R6}
\end{equation}
where $\alpha_{k_0E}$ and $\beta_{k_0E}$ are known as Bogoluibov coefficients. Since $u_E^*(x)$ corresponds to negative energy, the non-vanishing value of $\beta_{k_0E}$ yields the mixing of positive and negative energies in $\Psi_{k_0}$. Whether $u_E(x)$ is thermal with respect to $\Psi_{k_0}(x)$ is determined by the value of $\beta_{k_0E}$ (see section $3.2$ of  \cite{BookA1} for the justification). Therefore we will now calculate this coefficient. 

Use of the normalization condition on $u_E(x)$ yields
\begin{equation}
\beta_{k_0E} = \int_0^\infty dx \Psi_{k_0}(x)u_E(x)~.
\label{R7}
\end{equation}
Substitution of the respective eigenstates in the above we obtain
\begin{eqnarray}
\beta_{k_0E} &=& \frac{N}{\sqrt{2\pi\hbar\kappa}} \int_0^\infty dx e^{ik_0x}(x)^{-\frac{1}{2} + \frac{iE}{\hbar\kappa}}
\nonumber
\\
&=& \frac{N}{\sqrt{2\pi\hbar\kappa}} e^{-(\frac{1}{2} + \frac{iE}{\hbar\kappa})\ln(-ik_0)}\Gamma\Big(\frac{1}{2} + \frac{iE}{\hbar\kappa}\Big)
\nonumber
\\
&=& \frac{N}{\sqrt{2\pi\hbar k_0\kappa}} e^{i(\frac{\pi}{4} - \frac{E\ln k_0}{\hbar\kappa})} e^{-\frac{\pi E}{2\hbar\kappa}}\Gamma\Big(\frac{1}{2} + \frac{iE}{\hbar\kappa}\Big)~.
\label{R8}
\end{eqnarray}
To evaluate the integration we used the following formula
\begin{equation}
\int_0^\infty dx ~x^{s-1}e^{-bx} = e^{-s\ln b}\Gamma(s)~,
\label{R9}
\end{equation}	
with $\textrm{Re}(b)>0$ and $\textrm{Re} (s)>0$. In order to apply the above formula we identified $s=(1/2) + i(E/\hbar\kappa)$ and chose $b=-ik_0 + \epsilon$ with $\epsilon>0$ to make the integration convergent. At the end $\epsilon\to 0$ limit has been taken. Then one finds
\begin{equation}
|\beta_{k_0E}|^2 = \frac{N^2}{\hbar k_0\kappa}\frac{1}{e^{\frac{2\pi E}{\hbar\kappa}}+1}~.
\label{R10}
\end{equation}
Non-vanishing of this value signifies that the mixing has happened and $u_E$ appears to be thermal with respect to $\Psi_{k_0}$. One may see that $u_E$ will feel a temperature given by Hawking expression $T=\hbar\kappa/2\pi$. It may be mentioned that (\ref{R10}) gives the probability of finding negative energy state $u_E^*$. 

Similarly one can find the probability of finding positive energy state $u_E$ by calculating $|\alpha_{k_0E}|^2$. This is found out to be as
\begin{eqnarray}
|\alpha_{k_0E}|^2 &=& \Big|\int_0^\infty dx \Psi_{k_0}(x)u_E^*(x)\Big|^2
\nonumber
\\
&=& \frac{N^2}{\hbar k_0\kappa} \frac{e^{\frac{2\pi E}{\hbar\kappa}}}{\frac{2\pi E}{\hbar\kappa}+1}~.
\label{R11}
\end{eqnarray}
The ratio $|\beta_{k_0E}|^2/|\alpha_{k_0E}|^2$ then provides the information about the temperature \cite{Keski-Vakkuri:1996wom} seen by the particle with Hamiltonian (\ref{1}) for energy $E$. This is given by
\begin{equation}
\frac{|\beta_{k_0E}|^2}{|\alpha_{k_0E}|^2} = e^{-\frac{2\pi E}{\hbar\kappa}}~.
\label{R12}
\end{equation}
Comparing with the Boltzmann factor one identifies the temperature as $T=\hbar\kappa/2\pi$. Note that this has been obtained with respect to state $\Psi_{k_0}$ and therefore the eigenstate of (\ref{1}) appears to be thermal when one compares with $\Psi_{k_0}$. Moreover if the mixing was not there then $\beta_{k_0E}$ must vanish and in that case the above ratio also vanishes. So here the thermalization is due to the occurrence of this mixing.

Here we found that even if individual eigenstates are not showing any thermal property within their own territory, whereas one state is appeared to be thermal in nature with respect to other state. Therefore we call this mechanism as relative thermalization. This is happening to eigenstates of (\ref{1}) which incorporates a random property and therefore we argue that such randomness (provided by horizon) may be the underlying reason for horizon appears to be thermal with respect to this particle.

\end{widetext}	
\end{document}